\documentclass[aip,pof,amsmath,amssymb,showpacs,preprint] {revtex4-1}

\usepackage{graphicx}
\usepackage{dcolumn}
\usepackage{bm}
\usepackage{ifthen}
\usepackage{psfrag}
\usepackage{booktabs}
\usepackage{array}

\usepackage{color}
\newcommand{\hilight}[1]{#1}

\usepackage{calc}
\usepackage{amsmath}
\usepackage{xcolor}
  
\newcommand\HLbox[1]{#1}

\newcommand{\diff}[2]{\frac{\partial #1}{\partial #2}}

\newcommand{\klamm}[1]{\left( #1 \right)}

\newcommand\Rey{{\textit{Re}}}  

\newcommand{\defi}{{\mathrel{\mathop:}=}}

\newcommand{\Offiziell}{1}
\newcommand{\Comment}[1]{ \ifthenelse{\Offiziell = 0}{{\bf #1}}{}}

\newcommand{\bb}{\begin{equation}}
\newcommand{\ee}{\end{equation}}
\newcommand{\ba}{\begin{eqnarray}}
\newcommand{\ea}{\end{eqnarray}}

\usepackage{natbib}

\begin{document}


\title{Symmetry Analysis in Linear Hydrodynamic Stability Theory:\\ Classical and New Modes in Linear Shear}

\author{Andreas Nold}
\email{andreas.nold09@imperial.ac.uk}
\affiliation{Department of Chemical Engineering, Imperial College London, London, SW7 2AZ, United Kingdom}

\author{Martin Oberlack}
\affiliation{Chair of Fluid Dynamics, Technische Universit\"at Darmstadt, 64287 Darmstadt, Germany } 
\affiliation{Center of Smart Interfaces, Technische Universit\"at Darmstadt, 64287 Darmstadt, Germany }
\affiliation{Graduate School of Computational Engineering, TU Darmstadt, 64293 Darmstadt,  Darmstadt, Deutschland}

\date{\today}

\begin{abstract}
We present a symmetry classification of the linearised
Navier-Stokes equations for a two-dimensional unbounded linear shear
flow of an incompressible fluid. The full set of symmetries is employed to systematically
derive invariant ansatz functions. The symmetry analysis grasps
three approaches. Two of them are existing ones, representing the classical normal
modes and the Kelvin modes, while the third is a novel approach and
leads to a new closed-form solution of traveling modes, showing
qualitatively different behaviour in energetics, shape and
kinematics when compared to the classical approaches. The last modes
are energy conserving in the inviscid case. They are localized in
the cross-stream direction and periodic in the
streamwise direction. As for the kinematics, they travel at constant
velocity in the cross-stream direction, whilst in the streamwise
direction they are accelerated by the base flow. In the viscous case, 
the modes break down due to damping of high wavenumber contributions.
\end{abstract}


\maketitle

\section{Introduction} \label{Intro}

 Flow stability theory deals with the breakdown of an ordered
laminar non-uniform flow and the onset of turbulent structures.
This transition to turbulence can be found in nature and industrial
applications. For example, when observing rivers, winds as well as
pipelines and bearings, we notice that accelerating a flow leads to
turbulent structures. Apart from its economic relevance, the study
of the precise location, time and form of the transition to
turbulence has fascinated generations of scientists since the famous
experiment of an unstable flow in a pipe by Reynolds in 1883. 
Due to its complexity, even canonical examples such as the
stability of a simple Couette flow between two infinite plates,
where one plate is moving steadily in one direction and the other
plate is fixed, have been a topic of debate for decades.

The classical approach to stability problems is the so-called normal
mode approach as derived by \citet{Orr1907}. It consists of periodic
modes traveling in the streamwise direction and has been applied to
the linear stability problem of a plane Couette flow by
\citet{Hopf1914},  \citet{Wasow1953}, \citet{Grohne:1954ys},
\citet{1974Reid} and others, yielding a decay of all modes for large
times. A second viable approach consists of using the modes
introduced by \citet{Kelvin1887}, with a time-dependent wavelength
in the cross-stream direction. Rosen \cite{Rosen:1971fk} reviewed this
approach and formulated a general solution for perturbations of a
plane Couette flow in the linear framework. The three-dimensional
Kelvin modes exhibit a period of modest algebraic/ transient growth
before entering a phase of exponential viscous decay.  Decay of all
perturbations in shear flows for large times was also shown by
\citet{Case1960} in a closed-form solution of the bounded inviscid
initial-value problem, yielding an algebraic decay of the
perturbation with $1/t$ if the initially introduced vorticity is
finite. In particular, \citet{Romanov:1973kx} showed that all
eigenvalues for a small enough perturbation are less than $-C/\Rey$,
where $C$ is a positive constant. Clearly, these results contradict
experiments, where transition to turbulence of a Couette flow is
observed at Reynolds numbers of around $350$
\citep{Lundbladh:1991uq,Tillmark:1992fk}.

It was in the 1990's that a novel approach revealed an
explanation for this apparent paradox. It was shown that non-uniform
flows are non-normal: they are spectrally stable, but perturbations
are able to gain the basic (shear) flow energy transiently and,
consequently, exhibit strong growth during a limited time interval
\citep{Trefethen:1993uq,Butler:1992kx,Reddy:1993kx,Gustavsson:1991vn}.
In particular, this is observed in the short-term behaviour of
perturbed flows. In the case of large enough initial perturbations,
the strong short-term non-normal growth allows for non-linear
effects to take place, which regenerate the transiently growing
perturbations. This positive feedback-loop allows for the onset of
turbulence and is usually denoted as bypass-transition
\citep{Grossmann:2000fk, Schmid:2006ys,Horton:2010uq}.

The aim of this work is to perform symmetry analysis of the linear
stability problem of an unbounded Couette flow and to show how
this mathematical tool can shed new light on long-standing and well-known
problems. A symmetry is a transformation which maps the solution
manifold of a differential equation onto itself. Special solutions
which do not change under a symmetry transformation are denoted as
invariant solutions, or, if scaling is part of the transformation,
as self-similar solutions. Generally, invariant solutions are a
powerful tool for the systematic development of ansatz functions for
solutions of partial differential equations \hilight{\citep{Bluman:2002,STEEB07,Bluman:2010}}. Especially in the area of fluid
mechanics, these have been applied successfully in various fields of application
\citep{Boisvert:1983zr,simonsen1997self,OBERLACK:2001ly,Oberlack:2001ve,avramenko,barenblatt,grebenev,Anco1997}.

We search for invariant solutions in the context of the stability
of an unbounded Couette flow and perform a symmetry classification
of the linearised Navier-Stokes equations for two-dimensional
perturbations. We show that the century-old normal mode approach and
the Kelvin mode approach both turn out to be among a larger class of
invariant solutions. In particular, the normal-mode approach is
obtained by a successive symmetry reduction with respect to space-
and time-\hilight{translation} symmetries together with a scaling symmetry. The
Kelvin mode approach is obtained similarly. It is invariant with
respect to a combination of a time-\hilight{translation} symmetry and one symmetry
which is due to the linearity of the base flow.

Strikingly, symmetry methods also allow for a third class of
invariant solutions so far not known to the authors. In the inviscid
case, we obtain a new closed form solution, exhibiting qualitatively
new behaviour in kinematics, energetics, and shape. In particular,
the modes consist of vortices traveling at constant speed in the
cross-stream direction and being accelerated by the linear shear
base flow in the streamwise direction. The closed form solution also
reveals a particular shape of the modes, which are
energy-conserving, decay in the cross-stream
direction and are periodic in the streamwise direction. In the viscous
case, invariance of these modes is lost. We present a closed-form
solution of the initial value problem. In agreement with
expectations from Kelvin mode theory, the energy of the modes decays
exponentially due to viscous damping effects.

We emphasize that the new invariant
function limits its application to two-dimensional settings, as three-dimensional 
effects are not taken into account here.
Due to the traveling in the
cross-stream direction the new invariant function also limits its
validity by a finite time interval (during which the solution
reaches a boundary). 
These properties might be given in wind shear or ocean flows.
Due to its peculiar behaviour and its analytical
simplicity, we believe that despite the limitations, the new
approach presented here adds a new perspective to the understanding
of shear flow dynamics. The feasibility of the found solution may
be confirmed numerically by imposing the  invariant function in a
plane Couette flow and following the dynamics by direct numerical
simulation.

In \S~\ref{sec:SymmetryAnalysis} and
\S~\ref{sec:SymmetriesCouetteFlow}, we give a brief overview of
symmetry methods and introduce a full symmetry-classification
for the linear stability analysis of a Couette flow in stream function formulation.
We then show how symmetry methods allow for a systematic derivation
of the normal mode approach and the Kelvin mode approach in
\S~\ref{sec:NormalModes} and \ref{sec:KelvinModes}. In the following, we present the new
invariant modes in \S~\ref{sec:NewSelf-SimilarModes}. In the
inviscid case, a closed-form solution of the modes is derived in
\S~\ref{sec:InviscidCase}, whereas the viscous case is studied in
\S~\ref{sec:ViscousCase}. We conclude in \S~\ref{sec:Conclusion}
with a summary of our results and discussion. In the Appendix,
linearly independent solutions to the viscous problem employing the
new invariant approach are presented.

\section{Symmetry Analysis \label{sec:SymmetryAnalysis}}

We consider an unbounded parallel two-dimensional shear flow
$\klamm{U(y),0}^{T}$ with a perturbation of the form
$\klamm{u(x,y,t),v(x,y,t)}^{T}$. Applying the curl on the
momentum-equations for the perturbations and introducing a stream
function $\psi(x,y,t)$ yields the following linearized fourth order
partial differential equation for the stream function:
\begin{align}
\frac{\partial}{\partial t} \Delta \psi + U \frac{\partial}{\partial x}\Delta \psi   - U'' \frac{\partial \psi}{\partial x}
= \nu \Delta \Delta \psi. \label{eq:StreamFunctionEqLin}
\end{align}
where $\nu$ is the kinematic viscosity and $\Delta$ is the Laplace operator.  A symmetry of this differential
equation is given by a \hilight{point transformation ${\bf T} = \klamm{\tilde x, \tilde y, \tilde t,\tilde \psi}$ with}
 \begin{alignat}{2}
 \HLbox{ \tilde x} &\HLbox{= \tilde x\klamm{x,y,t,\psi;\varepsilon},}\qquad&
  \HLbox{\tilde y} &\HLbox{= \tilde y\klamm{x,y,t,\psi;\varepsilon},} \label{eq:ChangeOfVariables1}  \\
   \HLbox{\tilde t} &\HLbox{= \tilde t\klamm{x,y,t,\psi;\varepsilon},}\qquad&
    \HLbox{\tilde \psi} &\HLbox{= \tilde \psi\klamm{x,y,t,\psi;\varepsilon},}\label{eq:ChangeOfVariables2}    
 \end{alignat}
for which the transformed
quantities satisfy the transformed differential Eq. (\ref{eq:StreamFunctionEqLin}), yielding
\begin{align}
\frac{\partial}{\partial \tilde t}\tilde \Delta \tilde \psi + U \frac{\partial}{\partial \tilde x}\tilde \Delta \tilde \psi   - U'' \frac{\partial \tilde  \psi}{\partial \tilde  x}
= \nu \tilde \Delta \tilde \Delta \tilde \psi. \label{eq:StreamFunctionEqLinTilde}
\end{align}
In (\ref{eq:ChangeOfVariables1})-(\ref{eq:ChangeOfVariables2}), $\varepsilon \in \mathbb{R}$ is the group
parameter of the transformation \hilight{and we assume that ${\bf T}$ is a} \hilight{smooth function of the parameter $\varepsilon$}. As an example, in the case of a space \hilight{translation} transformation, $\varepsilon$ is equivalent to the
actual \hilight{translation} performed. The rate of change with which the variables are
transformed is then given by the tangent vector field \hilight{$\klamm{\xi^{x},\xi^{y},\xi^{t},\eta}$ }
of the map ${\bf T}$ at $\varepsilon = 0$, defined by
\begin{align}
\HLbox{\xi^{\{x,y,t\}}} = \left. \diff{T_{\{x,y,t\}} }{\varepsilon} \right|_{\varepsilon= 0}
\qquad\text{and}\quad
\eta=  \left. \diff{T_{\psi}}{\varepsilon} \right|_{\varepsilon= 0}.
\end{align}
Sophus Lie first introduced a special kind of transformations, the so-called Lie-point symmetries. 
In this case, the map ${\bf T}$ has the special property
of being uniquely defined by its tangent vector field, which can also be written as an infinitesimal generator
\begin{align}
 X \defi \HLbox{\xi^{x}} \diff{}{x} +\HLbox{\xi^{y}} \diff{}{y} +\HLbox{\xi^{t}} \diff{}{t}  + \eta \diff{}{\psi},
\end{align}
\hilight{and which forms a Lie-symmetry group through}
\begin{align}
\HLbox{{\bf T} = e^{\varepsilon X}{\bf x},}
\end{align}
\hilight{where ${\bf x} = \klamm{x,y,t,\psi}$ \cite{Bluman:2010}}.

The first powerful tool of symmetry analysis is the concept of invariant solutions.
In a nutshell, invariance means that a solution $\psi(x,y,t)$ is not changed 
by the application of the transformation ${\bf T}$. 
The mathematical condition for this is that \hilight{the solution ${\psi} =
{\psi}\klamm{x,y,t}$} does not change its functional form after
application of the infinitesimal generator:
\begin{align}
\left.X\klamm{{\psi} - {\psi}\klamm{x,y,t}} \right|_{ {\psi} = \psi\klamm{x,y,t}} =  \eta - \HLbox{\xi^{x}} \diff{\psi}{x} - \HLbox{\xi^{y}} \diff{\psi}{y} - \HLbox{\xi^{t}} \diff{\psi}{t} =0. \label{eq:SelfSimilarSol}
\end{align}
As a simple example, a solution which is invariant with respect to the
translational symmetry in $x$ is represented by $X = \frac{\partial}{\partial x}$. In this case, condition (\ref{eq:SelfSimilarSol}) yields
\begin{align}
  \diff{\psi}{x}  =0
\end{align}
and therefore the respective invariant solution does not depend on $x$.

Next to invariance, the second crucial tool of symmetry analysis is
the combination of different symmetries. Mathematically, two symmetries $X_1,X_2$ are combined by
superposing their infinitesimals $X \defi a_{1} X_{1} + a_{2}X_{2}$ with
\begin{align}
\HLbox{\xi^{\{x,y,t\}}} = a_{1} \HLbox{\xi^{\{x,y,t\},1}} + a_{2}\HLbox{\xi^{\{x,y,t\},2}}
\qquad \text{and}\qquad
\eta = a_{1}\HLbox{\eta^{1}} + a_{2}\HLbox{\eta^{2}}
\end{align}
and coefficients $a_{1,2}$.
For example, instead of working with a
solution which is invariant with respect to the space \hilight{translation} symmetry alone,
we can search for solutions which are invariant to a combined space- and 
time \hilight{translation} symmetry $ X = a_{1}X_{1}+a_{2}X_{2}$, such that (\ref{eq:SelfSimilarSol}) transforms to 
\begin{align}
 - a_{1} \diff{\psi}{x} - a_{2} \diff{\psi}{t} =0.
\end{align}
The corresponding invariant solution represents
a traveling wave $\psi\klamm{a_2 x - a_1 t}$ for which the velocity of propagation depends on the ratio of the coefficients $a_{1}$ and $a_{2}$. 

Together, the concept of invariance and of superposition of symmetries will
help to gain new insights for the systematic analysis of
ansatz functions for linear stability theory. In this work, we will show how 
normal modes, Kelvin modes, as well
as a new type of base solutions can be systematically derived by
searching for invariant solutions with respect to a combination of the full set of available symmetries.
 For a more detailed introduction
to Lie Symmetries, see \hilight{\citet{Bluman89,Bluman:2002,Bluman:2010},} Cantwell \cite{Cantwell} and \citet{STEEB07}. The
symmetries presented in this work were derived by means of the
Lie-Algorithm, using the GeM package of \citet{CH07} and the DESOLVE
package of \citet{CAVU00}.

\section{Symmetries of a planar Couette Flow \label{sec:SymmetriesCouetteFlow}}

In this section, we present a full symmetry classification of the problem in stream function formulation.
We have also performed a symmetry classification of the momentum equations and the continuity
equation. However, no additional symmetries exist for this formulation of the problem.
Consequently, we will present all results in the stream function
formulation.

\begin{table}
\begin{center}
\setlength{\extrarowheight}{2pt}
\begin{tabular}{l@{\hspace{0.7cm}} llll @{\hspace{0.7cm}} l@{\hspace{0.7cm}}  m{3.5cm}     }
\hilight{Generator} &  \multicolumn{4}{c}{\hilight{Infinitesimals}}  &Transformation & Free Parameter/\\
& \hilight{$\xi^{x,i}$} & \hilight{$\xi^{y,i}$}&\hilight{$\xi^{t,i}$} & \hilight{$\eta^{i}$}  &&Free Function \\\hline
$X_0 = \psi_{0}\diff{}{\psi}$& 0&0&0& $\psi_{0}$
  & $\tilde{\psi} = \psi + C\psi_{0}(x,y,t)$& $C\in \mathbb{R}$, $\psi_{0}$ solves (\ref{eq:StreamFunctionEqLin})\\
$X_1  = \diff{}{x}$ & $1$&$0$&$0$& $0$&  $\tilde{x} = x + x_0$ & $x_{0} \in \mathbb{R}$\\
$X_2 = \diff{}{t}$ & $0$&$0$&$1$& $0$ &    $\tilde{t} = t + t_0$ & $t_{0} \in \mathbb{R}$ \\
$X_3 = \psi \diff{}{\psi}$ & $0$&$0$&$0$& $\psi$&  $\tilde{\psi} = C \psi$ & $C \in \mathbb{R}$\\\hline
 $X_4  = At\diff{}{x} + \diff{}{y}$
 & $At$&$1$&$0$& $0$& $\tilde{x} = x - y_{0}At,$
& $y_{0} \in \mathbb{R}$  \\
&&&&&$\tilde{y} = y-y_{0}$
\end{tabular}
\end{center}
\caption{\hilight{Complete set of symmetries of Eq. (\ref{eq:StreamFunctionEqLin}) for two-dimensional perturbations of a viscous}\\
\hilight{fluid for general shear flows $U(y)$ and for linear shear flows $U(y) = Ay$.}
\hilight{$X_{0,1,2,3}$}:  Symmetries for an arbitrary parallel shear flow $U(y)$. $X_{4}$: Additional symmetry if the base flow is restricted to $U(y) = Ay$.
$X_{1}$ and $X_{2}$ are the space and time
\hilight{translation} symmetries. $X_{0}$ is the superposition symmetry and $X_{3}$ is the scaling symmetry. $X_{0}$ and $X_{3}$ are both due to the linearisation of the Navier-Stokes equations for small perturbations.
}
\label{tab:SymmLNSE}
\end{table}%

\hilight{We have performed a symmetry analysis for two separate cases: First, a general shear}
\hilight{flow $U(y)$ was considered. In this case, the stream function formulation of the problem (\ref{eq:StreamFunctionEqLin})}
\hilight{allows for four different symmetry transformations (see also table \ref{tab:SymmLNSE}).} \\
\hilight{Second, a linear shear flow $U(y) = Ay$ was considered. This restriction allows for an} \\
\hilight{additional symmetry which reflects the effects of a base flow with a constant shear}\\
\hilight{rate $A$ (see table \ref{tab:SymmLNSE}).}
As mentioned in section \ref{sec:SymmetryAnalysis},
a linear combination of the infinitesimals allows for the construction of insightful invariant solutions. A general symmetry can be defined by
\begin{align}
X = a_{1} X_{1} + a_{2} X_{2} + a_{3} X_{3} + a_{4} X_{4}. \label{eq:LinearCombination}
\end{align}
We conclude that finding the invariant solutions amounts to solving (\ref{eq:SelfSimilarSol}) with
\begin{align}
\HLbox{\xi^{\{x,y,t\}}} = \sum_{i=1}^4 a_{i} \HLbox{\xi^{\{x,y,t\},i}}
\qquad \text{and} \qquad
\eta = \sum_{i=1}^{4} a_{i} \HLbox{\eta^{i}},
\end{align}
where \hilight{$\xi^{\{x,y,t\},i}$ and $\eta^{i}$} are defined in table
\ref{tab:SymmLNSE}. A brief examination of Eq.
(\ref{eq:SelfSimilarSol}) shows that the resulting invariant
solutions remain unchanged up to an equal scaling of all parameters
by some factor $C\in \mathbb{C}$. This allows us to set one
parameter $a_{i}$ which is unequal zero to one, without loss of any
information. We are also not interested in solutions which are
invariant with respect to only one symmetry, because this leads to
trivial simplifications. Furthermore, we will exclude the
superposition symmetry $X_{0}$ from our further considerations, as
the symmetry itself already includes a solution of the equation
under consideration.

\begin{figure}
\begin{center}
\includegraphics{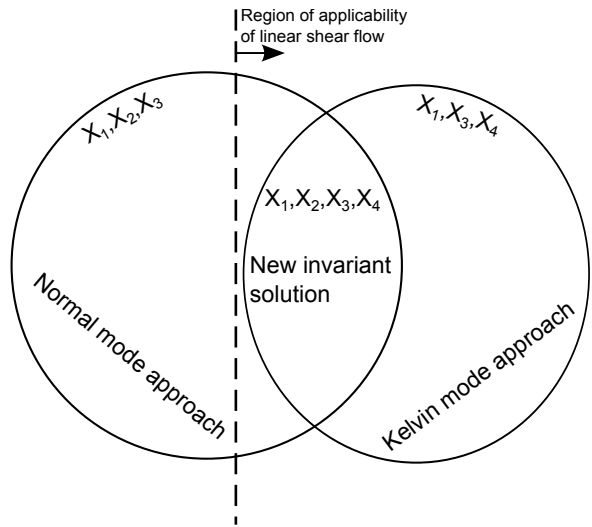}
\caption{Schematic view of selected invariant solutions. The symmetry $X_{4}$ is only valid for the case of a linear shear flow. If $X_{4}$ is not used, then successive symmetry reductions lead to the normal mode approach. Employing $X_{4}$, but excluding the time-\hilight{translation} symmetry $X_2$ leads to the Kelvin modes, whereas using the full set of symmetries leads to the new invariant solutions presented in this paper.}
\label{fig:DiffSymSolutions}
\end{center}
\end{figure}

All other invariant solutions can be divided into three classes,
such as shown in table \ref{tab:SymmetryCombination} and in Figure
\ref{fig:DiffSymSolutions}. In the following chapters we will
systematically derive the invariant solutions for these three
classes. For simplicity, we will present all solutions in
complex-valued form. As Eq. (\ref{eq:StreamFunctionEqLin}) is real-valued, the complex conjugate 
$\bar \psi$ of any solution $\psi$ will equally solve the problem. This allows us to 
construct a real-valued solution from any complex-valued $\psi$ by
 the simple superposition $\psi + \bar \psi$.

\begin{table}
\begin{center}
\begin{tabular}{l@{\hspace{0.5cm}}l@{\hspace{0.5cm}}l@{\hspace{0.5cm}}|l@{\hspace{0.5cm}}| l}
$a_{1}$ & $a_{2}$ & $a_{3}$ & $a_{4}$ &\\\hline
$\mathbb{C}$ & $\mathbb{C}$ & $\mathbb{C}$ & 0& Normal Modes\\
$\mathbb{C}$ & $0$ & $\mathbb{C}$ & 1 & Kelvin Modes\\
$\mathbb{C}$ & $1$ & $\mathbb{C}$ & $\mathbb{C}\backslash \{0\}$ & New invariant Modes\\
\end{tabular}
\end{center}
\caption{Classes of invariant solutions depending on the choice of parameters $a_{i}$. $a_{4}$ can only be nonzero in the case of a linear shear flow $U(y) =Ay$.
We note that the invariant solutions remain unchanged if the parameters $a_{i}$ are equally scaled by some coefficient $C\in \mathbb{C}$ (see Eq. \ref{eq:SelfSimilarSol}).}
\label{tab:SymmetryCombination}
\end{table}%

\section{Normal Modes \label{sec:NormalModes}}

The classical normal mode approach turns out to be an invariant solution with respect to the combination of the three symmetries $X_{1}$-$X_{3}$:
\begin{align}
\HLbox{X^{(\mathrm{N})}} \defi a_{1} \diff{}{x} + a_{2} \diff{}{t} + a_{3} \psi  \diff{}{\psi},
\end{align}
with the complex prefactors $a_{1,2,3} \in \mathbb{C}$ and $a_{4} = 0$ in (\ref{eq:LinearCombination}). Condition (\ref{eq:SelfSimilarSol}) then reduces to
\begin{align}
a_{3}\psi - a_{1} \diff{\psi}{x} - a_{2} \diff{\psi}{t} = 0. \label{eq:NormalModeCondEq}
\end{align}
 If $a_{2}\neq 0$, then the method of characteristics provides us with the solution
\begin{align}
\HLbox{\psi^{(\mathrm{N})}} (x,y,t) = \HLbox{f^{(\mathrm{N})}} \klamm{ \xi,y}e^{ \frac{a_{3}}{a_{2}} t}, \label{eq:InvariantSol_Orr}
\end{align}
with the new variable
\begin{align}
\xi = x-\frac{a_{1}}{a_{2}} t,
\end{align}
and where \hilight{$f^{(\mathrm{N})}$} solves the fourth order differential equation
\begin{align}
\klamm{U- \frac{a_{1}}{a_{2}}} \frac{\partial}{\partial \xi} \Delta \HLbox{f^{(\mathrm{N})}} + \frac{a_{3}}{a_{2}} \Delta \HLbox{f^{(\mathrm{N})}} - U'' \frac{\partial}{\partial \xi} \HLbox{f^{(\mathrm{N})}} = \nu \Delta \Delta  \HLbox{f^{(\mathrm{N})}} \label{eq:OrrSommerfeldEq1},
\end{align}
obtained by inserting ansatz (\ref{eq:InvariantSol_Orr}) into Eq. (\ref{eq:StreamFunctionEqLin}).
We note that this first symmetry reduction has not simplified the PDE considerably. While the number of
variables has been reduced by one, the PDE is still of fourth order and the number of parameters even increased by two: Additional to the viscosity $\nu$, we now also have $a_{1}/a_{2}$ and $a_{3}/a_{2}$. Fortunately, \hilight{by performing a symmetry analysis we find that Eq. (\ref{eq:OrrSommerfeldEq1})
admits two symmetries}: the scaling symmetry $f \diff{}{f}$ as well as the translational symmetry $\diff{}{\xi}$. Consequently, we repeat the procedure leading to (\ref{eq:InvariantSol_Orr}) by choosing $f$ to be invariant under a combination of both infinitesimal generators
\begin{align}
\HLbox{\tilde X^{(\mathrm{N})}} = b_{1} \diff{}{\xi} + b_{2} f \diff{}{f}, \label{eq:SecondSymmetryModal}
\end{align}
with the complex prefactors $b_{1,2} \in \mathbb{C}$.
Applying condition (\ref{eq:SelfSimilarSol}) for an invariant solution with respect to the general symmetry (\ref{eq:SecondSymmetryModal}), we obtain ansatz
\begin{align}
\HLbox{f^{(\mathrm{N}) }}(\xi, y) = \HLbox{{g}^{ (\mathrm{N})}}(y) e^{\frac{b_{2}}{b_{1}} \xi}.
\end{align}
Note that here we have assumed that $b_{1}\neq0$. If $b_{1} = 0$, then (\ref{eq:SelfSimilarSol}) leads to the trivial solution $\psi = 0$.
Inserting this approach into (\ref{eq:InvariantSol_Orr}) leads to
\begin{align}
\HLbox{\psi^{(\mathrm{N})}}(x,y,t) = \HLbox{g^{(\mathrm{N})}}(y) \exp\klamm{  \frac{b_{2}}{b_{1}}   x  + \frac{ a_{3} b_{1} - a_{1} b_{2} }{a_{2} b_{1}} t   }.
\end{align}
Let us now substitute the coefficients and assume that that the solution is bounded for $x\to \pm \infty$ (i.e. $\Re\klamm{b_{2}/b_{1}}= 0$). This yields the classical normal mode approach
\begin{align}
\HLbox{\psi^{(\mathrm{N})}}(x,y,t) = \HLbox{g^{(\mathrm{N})}}(y) e^{i{\alpha}(x - ct)}, \label{eq:AnsatzOSE}
\end{align}
with wavelength $\alpha = \Im \klamm{b_{2}/b_{1}}$ and wave speed $c = {a_1}/{a_2} - \klamm{a_3 b_1}/\klamm{a_2 b_2}$.
Insertion into (\ref{eq:StreamFunctionEqLin}) leads to the Orr-Sommerfeld equation
\begin{align}
\klamm{U-c}\klamm{\frac{d^2}{dy^2}-{\alpha}^2} \HLbox{g^{(\mathrm{N})} }&- U'' \HLbox{g^{(\mathrm{N})}}
= \frac{\nu}{i{\alpha} }\klamm{\frac{d^2}{dy^2}-{\alpha}^2}^2 \HLbox{g^{(\mathrm{N})}}.
\label{eq:OSE}
\end{align}
We note that the analysis can be repeated analogously if $a_2 = 0$ and $a_1 \neq 0$.
Summarizing this section, the Orr-Sommerfeld equation is derived through a successive symmetry reduction of the linearised Navier-Stokes equations each time using the full set of admitted symmetries for an arbitrary $U(y)$ in Eq. (\ref{eq:StreamFunctionEqLin}). This holds true for both the viscous and the inviscid case and is usually referred to as normal mode or modal approach. In this work, we will repeatedly apply the method of successive symmetry reductions in order to reproduce existing and find new approaches.

\section{Kelvin Modes \label{sec:KelvinModes}}
In the following, we restrict ourselves to the analysis of a linear shear flow
\begin{align}
U(y) = Ay.
\end{align}
Analogously to the derivation leading to the Orr-Sommerfeld equation, we search for a solution which is invariant with respect to the following combination of symmetries $X_{1}, X_{3}$ and $X_{4}$, excluding the time-\hilight{translation} symmetry by setting $a_{2} = 0$ in (\ref{eq:LinearCombination}):
\begin{align}
\HLbox{X^{(\mathrm{K})}}= a_{1} \diff{}{x} +a_{3} \psi\diff{}{\psi} + \klamm{At\diff{}{x} + \diff{}{y} }.
\end{align}
In this case, the defining Eq. (\ref{eq:SelfSimilarSol}) for the invariant solution becomes
\begin{align}
a_{3} \psi - \klamm{ a_{1} +At} \diff{\psi}{x} - \diff{\psi}{y} = 0.
\end{align}
The respective invariant solution is given by
\begin{align}
\HLbox{\psi^{(\mathrm{K})}}(x,y,t) = \HLbox{f^{(\mathrm{K})}}\klamm{\zeta, t} e^{a_{3} y}. \label{eq:KelvinSym1Ans}
\end{align}
with the new variable
\begin{align}
\zeta = x - y\klamm{At+a_{1}}.
\end{align}
\hilight{$f^{(\mathrm{K})}$} solves the equation obtained by inserting ansatz (\ref{eq:KelvinSym1Ans}) into Eq. (\ref{eq:StreamFunctionEqLin}):
\begin{align}
 \frac{\partial}{\partial t}  &\left(  \left(a_{3} - \klamm{At+a_{1}} \frac{\partial}{\partial \zeta}\right)^2 + \frac{\partial^2}{\partial \zeta^2}\right) \HLbox{f^{(\mathrm{K})}} =
 \nu \left(  \left( a_{3} - \klamm{ At +a_{1}}\frac{\partial}{\partial \zeta}\right)^2 + \frac{\partial^2}{\partial \zeta^2}\right)^{2}  \HLbox{f^{(\mathrm{K})}} .
\end{align}
Similar to the previous section, this equation admits the scaling symmetry $f \diff{}{f}$ and the translational symmetry $\diff{}{ \zeta}$. Superposing the two infinitesimal generators
\begin{align}
\HLbox{\tilde X^{(\mathrm{K})}} = b_{1}\diff{}{\zeta}+  b_{2} f \diff{}{f}, \label{eq:SecondSymmetryKelvin}
\end{align}
with the complex prefactors $b_{1,2} \in \mathbb{C}$ and
applying condition (\ref{eq:SelfSimilarSol}) for an invariant solution with respect to the general symmetry (\ref{eq:SecondSymmetryKelvin}), we obtain ansatz
\begin{align}
\HLbox{f^{(\mathrm{K})}}(\zeta, t) = \HLbox{{g}^{(\mathrm{K})}}(t) e^{\frac{b_{2}}{b_{1}} \zeta}. \label{eq:fk_Kelvin}
\end{align}
Note that here we have assumed that $b_{1}\neq0$. ($b_1 = 0$ only leads to the trivial solution $\psi =0$).
Inserting (\ref{eq:fk_Kelvin}) into (\ref{eq:KelvinSym1Ans}) leads to 
\begin{align}
\HLbox{\psi^{(\mathrm{N})}}(x,y,t) = \HLbox{g^{(\mathrm{N})}}(t) \exp\klamm{  \frac{b_{2}}{b_{1}}   \klamm{ x - y\klamm{At+a_{1}} } + a_{3}y   } .
\end{align}
Let us now substitute the coefficients $a_{1},a_{3},b_{1},b_{2}$, assuming that that the solution is bounded for $x\to \pm \infty$ (i.e. $\Re\klamm{b_{2}/b_{1}}= 0$) and for $y\to \pm \infty$ (i.e. $\Re\klamm{a_{3}}=0$). We also set $a_{1}$ to zero, as $a_{1}$ only leads to a time-\hilight{translation} of the solutions.
This yields the Kelvin mode approach
\begin{align}
\HLbox{\psi^{(\mathrm{K})}} (x,y,t) = \HLbox{g^{(\mathrm{K})}}(t) e^{i \kappa_{x} \klamm{x-Ayt} + i\kappa_{y} y}, \label{eq:Ansatz123}
\end{align}
with wavelength $\kappa_{x} \in \mathbb{R}$ in the streamwise direction and a time-dependent wavelength
in the cross-stream direction $\kappa_{y}-\kappa_{x} At$, where $\kappa_{y} \in \mathbb{R}$. Finally, inserting
the Kelvin mode approach into the stream function form of the Navier-Stokes equation (\ref{eq:StreamFunctionEqLin}) gives the following ODE for \hilight{$g^{(\mathrm{K})}$}:
\begin{align}
- \frac{d}{dt} \klamm{\left( \kappa_{x}^2 + (A\kappa_{x} t - \kappa_{y})^2 \right) \HLbox{g^{(\mathrm{K})}}}
= \nu \left(\kappa_{x}^2 + (A\kappa_{x} t - \kappa_{y})^2\right)^2 \HLbox{g^{(\mathrm{K})}}.
\end{align}
The solution of this first order ODE yields for the stream function is
\begin{align}
\HLbox{\psi^{(\mathrm{K})}}(x,y,t) =&\frac{\kappa_x^2 +
\kappa_y^2}{\kappa_x^2 + (\kappa_x  At - \kappa_y)^2} \exp\klamm{
i\kappa_x(x-Ayt)+i\kappa_y y } \times\notag
\\
&\times\exp\klamm{-\nu t\left(\frac{1}{3}\kappa_x^2A^{2}t^2 -\kappa_y\kappa_x At +\kappa_y^2 + \kappa_x^2 \right)}, \label{eq:KelvinsSolutions}
\end{align}
which corresponds to the solution derived by \citet{Rosen:1971fk}, for two dimensional perturbations.
Here, we see that the linear shearing and with that the time-dependent wavelength in cross-stream direction in fact reflects the effect of the symmetry $X_{4}$ of the base flow.
However, we emphasise that in the derivation of the Kelvin modes, not the full set of symmetries was used. Instead, the time \hilight{translation} symmetry $X_{2}$, i.e. $\tilde t =t + t_{0}$ in Table \ref{tab:SymmLNSE}, was left out by setting the corresponding group parameter $a_{2}$ artificially to zero.

\section{New Invariant Modes \label{sec:NewSelf-SimilarModes} }

The symmetry analysis leading to the Kelvin modes excluded the time \hilight{translation} symmetry $X_{2}$, i.e. $\tilde t = t+t_{0}$ (see also table \ref{tab:SymmLNSE}). We will now show how a new class of ansatz functions can be obtained by also including this symmetry. The general infinitesimal in this case is (\ref{eq:LinearCombination}) with $a_{2}$ and $a_{4}$ nonzero. As described in section \ref{sec:SymmetryAnalysis}, we can rescale the infinitesimal generator by an arbitrary constant. For means of simplicity, we rescale the infinitesimal generator such that $a_{2} = 1$:
\begin{align}
\HLbox{X^{(\mathrm{I})}} = a_{1} \diff{}{x} + \diff{}{t} + a_{3} \psi\diff{}{\psi} + a_{4} \klamm{At\diff{}{x} + \diff{}{y} }.\label{eq:NewSelfSimilarSymmetry1}
\end{align}
Condition (\ref{eq:SelfSimilarSol}) then becomes
\begin{align}
a_{3} \psi - \klamm{ a_{1} +a_{4} At} \diff{\psi}{x} - a_{4}\diff{\psi}{y}  - \diff{\psi}{t}= 0,
\end{align}
which is solved by
\begin{align}
\HLbox{\psi^{(\mathrm{I})}} (x,y,t) = \HLbox{f^{(\mathrm{I})}}\klamm{ \bar x, \bar y} e^{ a_{3} \klamm{ t  + \frac{a_{1}}{Aa_{4}} } }, \label{eq:PModesFirstAnsatz}
\end{align}
with the new variables
\begin{align}
\bar x =   x-  \frac{Aa_{4}}{2} \klamm{ t + \frac{a_{1}}{aa_{4}} }^{2}
\qquad  \text{and}\qquad
\bar y =  y- a_{4}\klamm{ t+ \frac{a_{1}}{Aa_{4}} } .  \label{eq:SelfSimilarVariablesNew}
\end{align}
Insertion into (\ref{eq:StreamFunctionEqLin}) yields
\begin{align}
 \left(  A \bar y  \frac{\partial}{\partial \bar x} - a_{4} \frac{\partial}{\partial \bar y} + a_{3} \right)\Delta f^{\HLbox{(\mathrm{I})}}  &  =\nu \Delta\Delta f^{\HLbox{(\mathrm{I})}} . \label{eq:Linear_Formula_1}
\end{align}
Analogously to the derivation of the normal and the Kelvin modes, we apply a successive symmetry reduction.
Equation (\ref{eq:Linear_Formula_1}) admits a scaling symmetry and a translational symmetry in $\bar x$, yielding the general infinitesimal generator
\begin{align}
\tilde X^{\HLbox{(\mathrm{I})}}  = b_{1}\diff{}{\bar x} + b_{2} f \diff{}{f}.
\end{align}
The invariant solution to this symmetry is given by
\begin{align}
f^{\HLbox{(\mathrm{I})}} (\bar x, \bar y) = g^{\HLbox{(\mathrm{I})}} \klamm{ \bar y} e^{ \frac{b_{2}}{b_{1}} \bar x},
\end{align}
which - after insertion into Eq. (\ref{eq:PModesFirstAnsatz}) - yields the new ansatz function
\begin{align}
\psi^{\HLbox{(\mathrm{I})}} (x,y,t) =& g^{\HLbox{(\mathrm{I})}} \klamm{  y- a_{4}\klamm{ t+ \frac{a_{1}}{Aa_{4}} } } \times\notag\\
&\times\exp\klamm{ \frac{b_{2}}{b_{1}}\klamm{ x-  \frac{Aa_{4}}{2} \klamm{ t + \frac{a_{1}}{Aa_{4}} }^{2}  }  + a_{3} \klamm{ t  + \frac{a_{1}}{Aa_{4}} }  }.
\end{align}

Let us now rename the coefficients, and assume that
$\Re\klamm{b_{2}/b_{1}} = 0$ in order to assure that the solution
remains bounded for $x \to \pm \infty$. We also set $a_{1}$ to zero,
as this parameter only leads to a time offset. Furthermore, we scale
the argument of $g^{\HLbox{(\mathrm{I})}} $ with $\kappa$ and obtain
\begin{align}
\psi^{\HLbox{(\mathrm{I})}} (x,y,t)& = g^{\HLbox{(\mathrm{I})}} \klamm{\kappa y - \frac{t}{T} } \exp\klamm{ i\kappa \klamm{x  - \frac{At^{2}}{2 \kappa T} } + c \kappa t }.
\label{eq:psis_S}
\end{align}
Based on the coefficients $a_{i}$ and $b_{i}$, we have introduced the wavelength $\kappa = \Im\klamm{\frac{b_{2}}{b_{1}}}$, the time-scale $T$, such that $\kappa T = \frac{1}{a_{4}}$. We also introduce a parameter $c = a_{3}\kappa^{-1}$. As explained above, $\kappa$ is real. In the next section, we will show that in the inviscid case, physical consistency requires $T$ to be real and $c$ to be imaginary. 

\begin{figure}
\begin{center}
\includegraphics{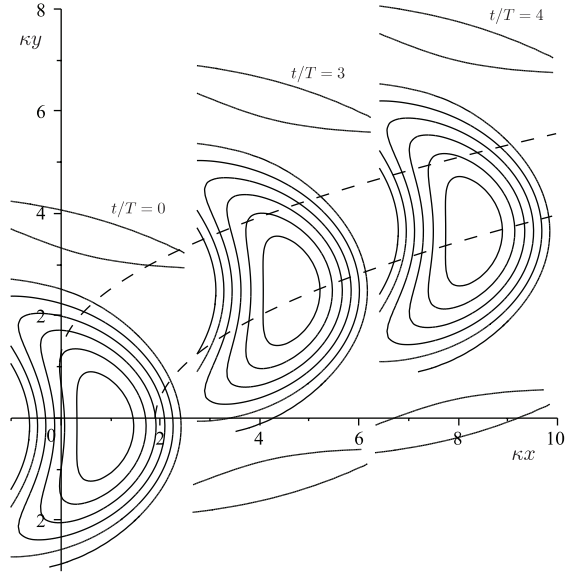}
\caption{Perturbation streamlines of the new inviscid invariant modes for a shear rate  $A = 1/T$ at different points in time. Note that at every instant in time, the solution is periodic in $x$. In time, the streamlines are translated along the dashed lines, defined by Eq. (\ref{eq:xtyt_NewSelfSimilarMode}).}
\label{fig:SelfSimilarModes}
\end{center}
\end{figure}

The novel ansatz function (\ref{eq:psis_S}) describes modes
traveling at a constant speed $\klamm{\kappa T}^{-1}$ in the
cross-stream direction. In the streamwise direction, the modes are
periodic and are accelerated by the base flow, such that for an
outer observer, the modes travel in parabola-shaped curves described
by
\begin{align}
\left( \begin{array}{c} x\klamm{t} - x_{0}  \\ y\klamm{t}  -y_{0}\end{array} \right) =
 \frac{1}{\kappa T}
\left( \begin{array}{c}
 \frac{1}{2}At^{2}  \\
t
\end{array} \right) .
\label{eq:xtyt_NewSelfSimilarMode}
\end{align}
Here, $(x_{0},y_{0}) \in \mathbb{R}^{2}$ defines the initial position of the mode (see Figure \ref{fig:SelfSimilarModes}) and $c=0$ for simplicity.
The shape of the modes is defined by $g^{I}$, which has to satisfy the equation
\begin{align}
\klamm{ -  \frac{d}{d \tilde y} + i S \tilde y  + \tilde c  }&\klamm{\frac{d^{2}}{d \tilde y^{2} } - 1}  g^{\HLbox{(\mathrm{I})}}  =
 \frac{1}{\Rey} \klamm{\frac{d^{2}}{d \tilde y^{2} } - 1}^{2}  g^{\HLbox{(\mathrm{I})}} ,
 \label{eq:SelfSimilarModes_1}
\end{align}
with
\begin{align}
\tilde y = \kappa y - \frac{t}{T},
\end{align}
obtained by insertion of (\ref{eq:psis_S}) into (\ref{eq:StreamFunctionEqLin}).
Note that we have formulated the problem in dimensionless form, introducing the dimensionless shear rate $S$, the growth rate $\tilde c$
and a Reynolds-number:
\begin{align}
S \defi AT \quad,\quad \tilde c \defi \kappa T c
\qquad\text{and}\qquad \Rey \defi \frac{1}{\nu \kappa^{2}T }.
\end{align}
In the sequel, we present an invariant solution of (\ref{eq:SelfSimilarModes_1}) in the inviscid case.
We note that in the viscous case, it is not possible to obtain \hilight{physically consistent} invariant solutions (see Appendix \ref{sec:AppendixViscousDivergingSolution} for details).

\subsection{The Inviscid Case \label{sec:InviscidCase}}

In this section, we give a detailed analysis of the invariant
modes (\ref{eq:psis_S}) in the inviscid case, including a proof of their energy-conservation and a study of
the vorticity- and the velocity field. We will also present a link to the Kelvin modes.

Integrating the first differential operator of Eq. (\ref{eq:SelfSimilarModes_1}) yields up to a constant pre factor
\begin{align}
\klamm{\frac{d^{2}}{d \tilde y^{2} } - 1}  g^{\HLbox{(\mathrm{I})}} _{\infty} = \exp\klamm{ \frac{iS\tilde y^{2}}{2} + \tilde c \tilde y}. \label{eq:InviscidSelfSEq1}
\end{align}
Note that according to ansatz (\ref{eq:psis_S}), up to a pre factor $\kappa^{2}$ the expression above corresponds with the negative vorticity for $x,t = 0$. We require that the initial vorticity remains finite for $y\to \pm \infty$, such that necessarily we have to require that $S$ is real (which implies that $T$ is real) and that $\Re\klamm{\tilde c} = 0$. The imaginary part of $\tilde c$ will only lead to a \hilight{translation} in the cross-stream direction, such that for simplicity we set $\tilde c = 0$.
Requiring $\Re\klamm{\tilde c} = 0$ enforces zero exponential growth or decay of the amplitude of the modes over
time. Solving Eq. (\ref{eq:InviscidSelfSEq1}) for
$g^{\HLbox{(\mathrm{I})}} _{\infty}$ then yields the closed-form analytical solution
\begin{align}
g^{\HLbox{(\mathrm{I})}} _{\infty}(\tilde y) =e^{\tilde y} \text{erfc}&\klamm{\frac{1 - i}{2}\sqrt{S}  \klamm{\tilde y+  \frac{i}{S} } }+
 e^{-\tilde y} \text{erfc}\klamm{\frac{-1 + i}{2} \sqrt{S} \klamm{\tilde y-\frac{i}{S} } } .
 \label{eq:SelfSimilarInviscidG}
\end{align}
In the following, we will scrutinize this result and show some of its physical implications. 

\begin{figure}
\begin{center}
\includegraphics{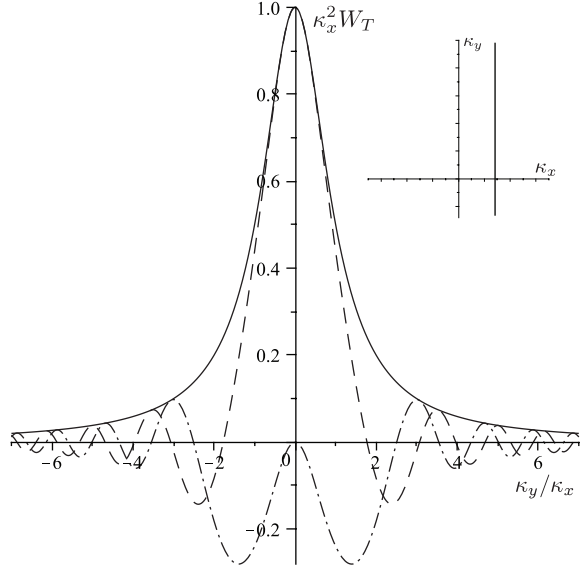}
\caption{
$\kappa_{y}$-dependence of the stream function of the new invariant solution in phase-space at time $t=0$ and for a fixed wavelength $\kappa_x = \kappa$ in $x$-direction (see $W_{T}$ in (\ref{eq:WAlpha})). The stream functions were normalized s.t. the maximal value corresponds to unity. The inset depicts the stream function in wave-space in a $\kappa_{x}$-$\kappa_{y}$ plane. Solid line: Absolute value $|W_{T}|$. Dashed line: Real part $\Re\klamm{W_{T}}$. Dash-Dotted Line: Imaginary part $\Im\klamm{W_{T}}$. 
}
\label{fig:FourierDecomposition}
\end{center}
\end{figure}

The peculiar
form of the invariant solution can be obtained by a special
combination of Kelvin modes (see Eq. (\ref{eq:KelvinsSolutions})). This combination can be obtained by decomposing the initial condition into Fourier modes
\begin{align*}
\psi^{\HLbox{(\mathrm{I})}} (x,y,0) = \iint W_{T}\klamm{\kappa_{x},\kappa_{y}}  e^{ i \kappa_{x} x + i \kappa_{y}y }  d\kappa_{x} d\kappa_{y},
\end{align*}
weighted with
\begin{align}
W_{T}\klamm{\kappa_{x},\kappa_{y}} &= \frac{ \exp\klamm{ - \frac{i}{2AT}  \frac{\kappa_{y}^{2}}{\kappa_{x}^{2}}} }{\kappa_{x}^{2}+ \kappa_{y}^{2}} \delta\klamm{\kappa_{x}-\kappa}. \label{eq:WAlpha}
\end{align}
For a comparison of the new invariant
solution with the Kelvin mode solution in phase space, see Figure
\ref{fig:FourierDecomposition}.
A quick computation confirms that
this is consistent with modes traveling in parabola-shaped curves:
\begin{align}
\psi^{\HLbox{(\mathrm{I})}} (x,y,t) &= \iint W_{T}\klamm{\kappa_{x},\kappa_{y}} \frac{\kappa_{x}^{2}+\kappa_{y}^{2}}{\kappa_{x}^{2} + \klamm{\kappa_{y} - \kappa_{x}At}^{2} }  e^{ i \kappa_{x}\klamm{ x-yAt} + i \kappa_{y}y }  d\kappa_{x} d\kappa_{y}\label{eq:psiSW}\\
&= \int
\frac{ e^{ - \frac{i}{2AT} \frac{\klamm{\bar \kappa_{y}+\kappa At}^{2}}{\kappa_{x}^{2}}} }{\kappa^{2}+ \bar \kappa_{y}^{2}}
 e^{ i \kappa x + i \bar \kappa_{y}y } d\bar\kappa_{y} \label{eq:psiSW2} \\
 &=
 \int  \frac{ e^{ - \frac{i}{2AT} \frac{\bar \kappa_{y}^{2}}{\kappa_{x}^{2}}} }{\kappa^{2}+ \bar \kappa_{y}^{2}}
 e^{ i \kappa\klamm{ x   - \frac{At^{2}}{2\kappa T}  } + i \bar \kappa_{y} \klamm{  y - \frac{t}{\kappa T} } }  d\bar \kappa_{y}\\
 &= \psi^{\HLbox{(\mathrm{I})}} \klamm{x - \frac{At^{2}}{2\kappa T} , y - \frac{t}{\kappa T},0}.
\end{align}

\begin{figure}
\begin{center}
\includegraphics{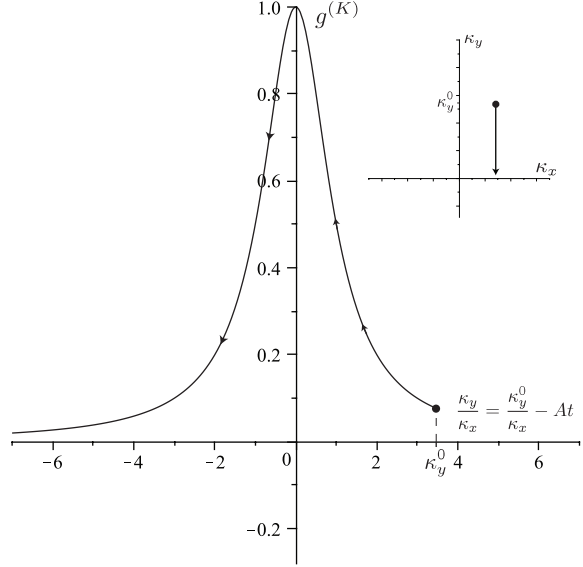}
\caption{
Time-Evolution of the stream function of a single Kelvin mode (\ref{eq:KelvinsSolutions}) in phase space.
The stream functions was normalized s.t. the maximal value corresponds to unity. The inset depicts the stream function in wave-space in a $\kappa_{x}$-$\kappa_{y}$ plane.
}
\label{fig:FourierDecomposition1}
\end{center}
\end{figure}

\begin{figure}
\begin{center}
\includegraphics{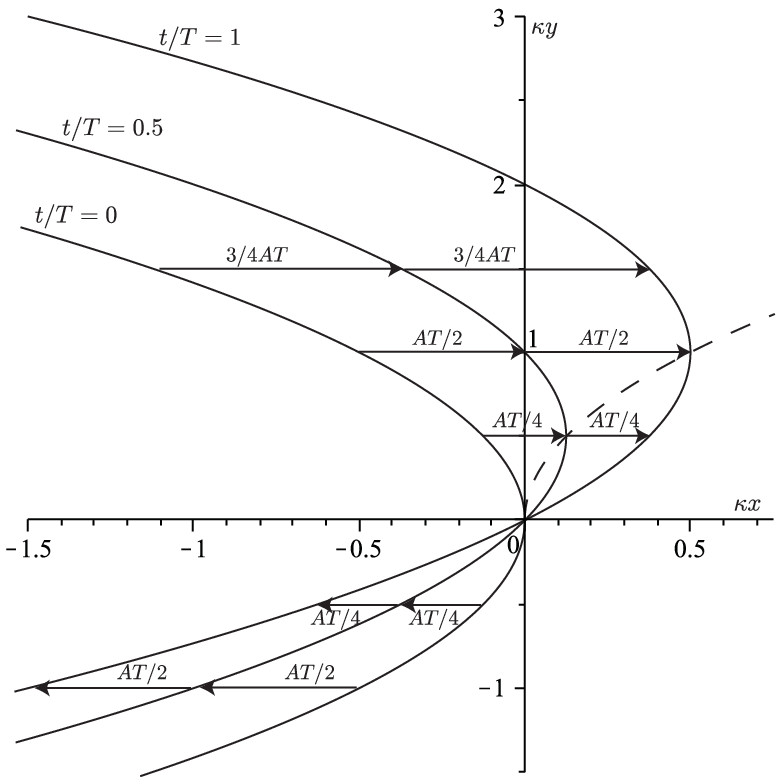}
\caption{Solid lines: Isoline of the vorticity at different points in time for shear rates $A = 1/T$. Following (\ref{eq:VorticitySol}), the isolines at time $t$ are described by $(\kappa x-\frac{At^{2}}{ 2 T }) + \frac{AT}{2}\klamm{ \kappa y - \frac{t}{T} }^{2}= c$ for some $c\in \mathbb{R}$. With time, shearing of the parabola-shaped isolines leads to a displacement following a parabola-shaped trajectory, described by $\kappa x =  \kappa x_{0} + \frac{At^{2}}{2T}$ and $\kappa y = \kappa y_{0} + \frac{t}{T}$ (dashed line).}
\label{fig:ParabolaVorticityIsolines}
\end{center}
\end{figure}

Let us now examine the vorticity of the solution. According to (\ref{eq:psis_S}) and (\ref{eq:InviscidSelfSEq1}), we obtain
\begin{align}
\omega(x,y,t) = - \Delta \psi &= -\kappa^{2}  \exp\klamm{ i\kappa \klamm{ x- \frac{At^{2}}{ 2\kappa T }  } + i \frac{AT}{2} \klamm{\kappa y - \frac{t}{T}}^{2}} \label{eq:VorticitySol}\\
&= -\kappa^{2}  \exp\klamm{ i\kappa \klamm{ x- Ayt  } + i \frac{AT}{2} \kappa^{2} y^{2} } \notag\\
&= \omega\klamm{x-Ayt, y, 0}.\notag
\end{align}
Interestingly, the parabola-shaped trajectory of the perturbations,
given in Eq. (\ref{eq:xtyt_NewSelfSimilarMode}), is consistent with
a shearing of the vorticity over time. This is due to the
parabola-shaped isolines of the vorticity (see Figure
\ref{fig:ParabolaVorticityIsolines}). In particular, the mapping
\begin{align}
h(x,y) \quad &\to \quad h\klamm{x- Ayt,y},
\end{align}
when applied on a simple parabola $h(x,y) = x+ by^{2}$, yields
\begin{align}
x+by^{2}  \quad  &\to  \quad x-Ayt +by^{2} = x- \frac{(At)^{2}}{4b} + b\klamm{y- \frac{At}{2b}}^{2},
\end{align}
where in this case $b = AT\kappa/2$. Consequently, shearing the parabola is equivalent to translating it in a parabolic trajectory. This property was automatically made use of by the application of symmetry methods.

\begin{figure}
\begin{center}
\includegraphics{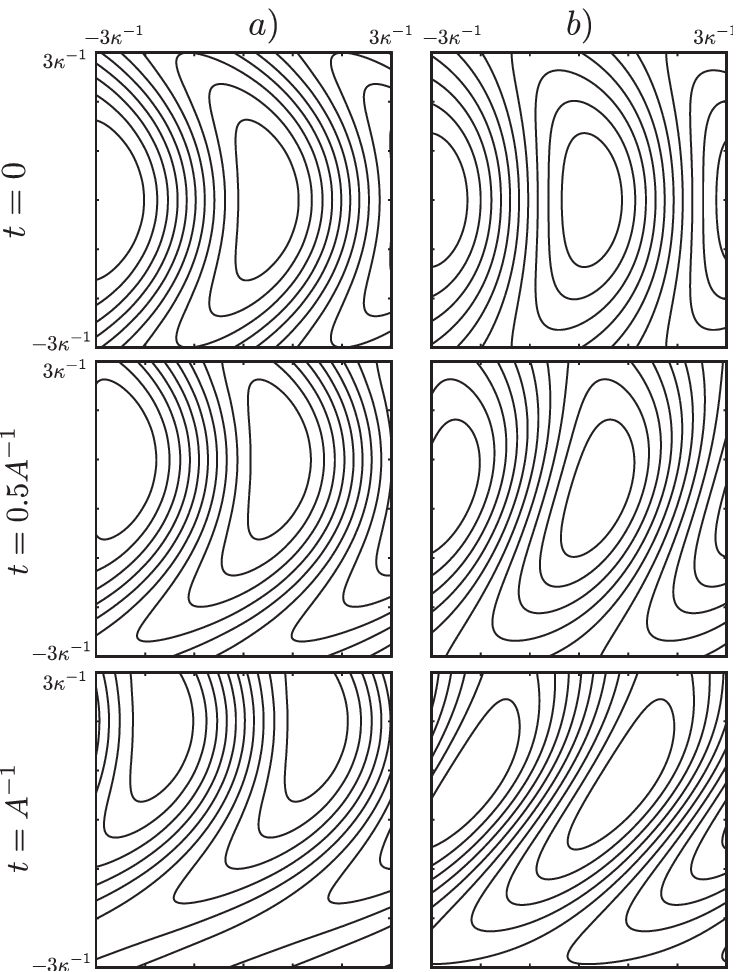}
\caption{Evolution of perturbation streamlines for two energy-conserving modes: a) the invariant modes (\ref{eq:psiSW}) obtained by symmetry-analysis and b) energy-conserving modes where $\tilde W$ (\ref{eq:WAlphaNew}) was used instead of (\ref{eq:WAlpha}). It can be observed how the invariant modes conserve their shape whereas the modes b) are distorted with time.}
\label{fig:CompareAlpha}
\end{center}
\end{figure}

Another interesting property of the new invariant modes is that they are energy-conserving. \hilight{This can be shown by integration in wave space,}
\begin{align}
\HLbox{ \iint |\hat\psi^{(\mathrm{I})}|^2 d\bar \kappa_y d \kappa_x = \iint \frac{1}{\klamm{\kappa^2 +  \bar \kappa_y^2}^{2}} \delta\klamm{\kappa_x - \kappa} d\bar \kappa_y d\kappa_x = \frac{ \pi }{2\kappa^{3}},}
\end{align}
\hilight{where $\hat \psi^{(\mathrm{I})}$ is the representation of $\psi^{(\mathrm{I})}$ in wave space, obtained from Eq.  (\ref{eq:psiSW2}).}
Interestingly, the modes presented here are not the only energy-conserving solutions of (\ref{eq:StreamFunctionEqLin}).
$W_{T}\klamm{\kappa_{x},\kappa_{y}}$ can be manipulated by changing its phase in order to obtain further
energy - conserving modes, e.g. by setting
\begin{align}
\tilde W_{T}\klamm{\kappa_{x},\kappa_{y}} &= \frac{ \exp\klamm{ - \frac{i}{2AT} \frac{\kappa_{y}^{4}}{\kappa_{x}^{2}}} }{\kappa_{x}^{2}+ \kappa_{y}^{2}} \delta\klamm{\kappa_{x}-\kappa}. \label{eq:WAlphaNew}
\end{align}
In Fig. \ref{fig:CompareAlpha}, we compare the time evolution of streamlines of the invariant modes (\ref{eq:WAlpha}) and (\ref{eq:WAlphaNew}). We note that the modes described by (\ref{eq:WAlphaNew}) are not invariant, do not conserve their shape and hence are of a different nature than the modes obtained by symmetry analysis.

\begin{figure}
\begin{center}
\includegraphics{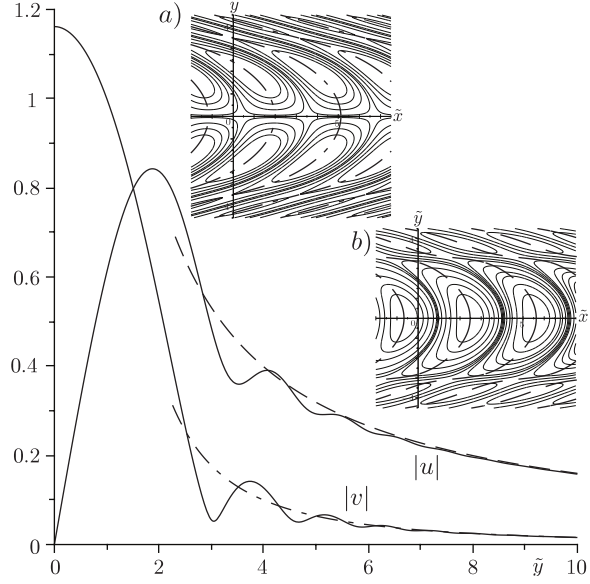}
\caption{Plot of the absolute value of  the velocity components $u$
and $v$ in the streamwise and the cross-section direction as
functions of $\tilde y =  \kappa  y - \frac{t}{T}$ for shear rates $A=1/T$. The dashed-dotted and dashed
lines correspond with the asymptotic behaviour given in Eqs.
(\ref{eq:AbsoluteAsymptoticBehavior}) and
(\ref{eq:AbsoluteAsymptoticBehaviorPrime}). The insets a) and b) show
isolines of the velocity components $u$ and $v$, respectively. Both are plotted in a
moving frame of reference, given by $\tilde x  = \kappa x -
\frac{At^{2}}{2T}$ and $\tilde y$ as above.}
\label{fig:SelfSimilarModes_gVsg2}
\end{center}
\end{figure}

Let us now study the implications for the far-field behavior of the velocity field of the perturbations. These can be written as
\begin{align}
u &=  \kappa \klamm{g_{\infty}^{\HLbox{(\mathrm{I})}} }' \klamm{\kappa y - \frac{t}{T}}  e^{i \klamm{\kappa x - \frac{At^{2}}{2T} }}\\
\qquad \text{and} \qquad
v &= -  i\kappa g_{\infty}^{\HLbox{(\mathrm{I})}} \klamm{\kappa y - \frac{t}{T}} e^{i \klamm{\kappa x - \frac{At^{2}}{2T} }}  .
\end{align}
Hence, in a moving frame of reference defined by $\tilde y = \kappa y - \frac{t}{T}$, the velocities decay algebraically as
\begin{align}
|u| &\sim (AT)^{-1/2} \tilde y^{-1} + O\klamm{\tilde y^{-3}}  \label{eq:AbsoluteAsymptoticBehavior}\\
\text{and} \qquad
|v| &\sim (AT)^{-3/2} \tilde y^{-2} + O\klamm{\tilde y^{-4}}  \label{eq:AbsoluteAsymptoticBehaviorPrime}
\end{align}
for $\tilde y \to \infty$. In Figure \ref{fig:SelfSimilarModes_gVsg2}, we compare the algebraic decay of the velocities in the cross-stream and the streamwise direction.

This leads to another major difference of the new invariant modes if compared to the Kelvin modes: Whereas the Kelvin modes 
represent a complete set of solutions to (\ref{eq:StreamFunctionEqLin}), the spectrum $W_{T}\klamm{\kappa_{x},\kappa_{y}}$ of the new invariant solutions is always even. A superposition of even functions will always be even, too. Hence, the set of solutions obtained by using the new invariant functions as basis functions is more restricted in that it only accepts solutions with an even spectrum in cross-stream direction. Following, we show how one can superimpose the new invariant solutions with a weight function $V_{\kappa_{Y}}\klamm{T} = -\frac{\kappa_{Y}}{2\pi AT^{2} \kappa_{x}^{2}} e^{\frac{i}{2AT} \frac{\kappa_{Y}^{2}}{\kappa^{2}}}$ to obtain the subset of even functions spanned by the Kelvin mode solutions. In particular, we show how to obtain the sum of two Kelvin mode solutions with $\kappa_y$ of different sign:

\begin{align}
\int \psi^{\HLbox{(\mathrm{I})}}  \klamm{x,y,t;T} V_{\kappa_{Y}}\klamm{T}  dT
=& \iint  \klamm{\int  W_{T}\klamm{\kappa_{x},\kappa_{y}} V_{\kappa_{Y}}\klamm{T}  dT} \times \notag\\
&\times\frac{\kappa_{x}^{2}+\kappa_{y}^{2}}{\kappa_{x}^{2} + \klamm{\kappa_{y} - \kappa_{x}At}^{2} }  e^{ i \kappa_{x}\klamm{ x-yAt} + i \kappa_{y}y }  d\kappa_{x} d\kappa_{y}\\
=& e^{ i \kappa\klamm{ x-yAt} } \klamm{ 
 \frac{e^{ i \kappa_{Y}y } }{\kappa^{2} + \klamm{\kappa_{Y} - \kappa At}^{2} } +
 \frac{e^{- i \kappa_{Y}y } }{\kappa^{2} + \klamm{\kappa_{Y} + \kappa At}^{2} }  },\\
=& \psi^{\HLbox{(\mathrm{K})}} \klamm{x,y,t;\kappa,\kappa_{Y}} + \psi^{\HLbox{(\mathrm{K})}} \klamm{x,y,t;\kappa,-\kappa_{Y}}
\end{align}
where we have used (\ref{eq:psiSW}) and that
\begin{align}
\int  W_{T}\klamm{\kappa_{x},\kappa_{y}} V_{\kappa_{Y}}\klamm{T} dT
&=\frac{\kappa_{Y}}{2\pi A \kappa_{x}^{2}}  \frac{\delta\klamm{\kappa_{x}-\kappa}}{\kappa_{x}^{2}+ \kappa_{y}^{2}} \int  \exp\klamm{  \frac{i}{2AT} \frac{\kappa_{Y}^{2} - \kappa_{y}^{2} }{\kappa_{x}^{2}}} d\klamm{\frac{1}{T}}  \\
&= 2\frac{ \kappa_{Y}}{\kappa_{x}^{2}+ \kappa_{y}^{2}} \delta\klamm{\kappa_{x}-\kappa} \delta\klamm{\kappa_{y}^{2}-\kappa_{Y}^{2}}\\
&=  \frac{\delta\klamm{\kappa_{x}-\kappa}}{\kappa_{x}^{2}+ \kappa_{y}^{2}}  \klamm{\delta\klamm{\kappa_{y}-\kappa_{Y}}+\delta\klamm{\kappa_{y}+\kappa_{Y}}}.
\end{align}
Contrary to the Kelvin modes, the new modes are also not
orthogonal, which is verified in phase space by:
\begin{align}
\int W_{T_{1}}\klamm{\kappa_{y}} \overline{ W_{T_{2}}\klamm{\kappa_{y}} } \klamm{\frac{{\kappa_{x}^{2}+\kappa_{y}^{2}}  }
{\kappa_{x}^{2}+\klamm{\kappa_{y} -\kappa_{x}A t}^{2}}}^{2} d\kappa_{y}
&= \int  \frac{ e^{- \frac{i}{2A}  ( \frac{1}{T_{1}} - \frac{1}{T_{2}})\frac{\kappa_{y}^{2}}{\kappa^{2}} } }{\klamm{\kappa_{x}^{2}+\klamm{\kappa_{y} -\kappa_{x}A t}^{2}}^2}   d\kappa_{y}.
\end{align}

We conclude that the new invariant modes form a non-orthogonal set of energy-conserving solutions which are even in cross-stream direction. Each invariant solution travels on a parabola-shaped curve with a constant velocity in cross-stream direction. 

\subsection{The Viscous Case \label{sec:ViscousCase}}
\begin{figure}
\begin{center}
\includegraphics{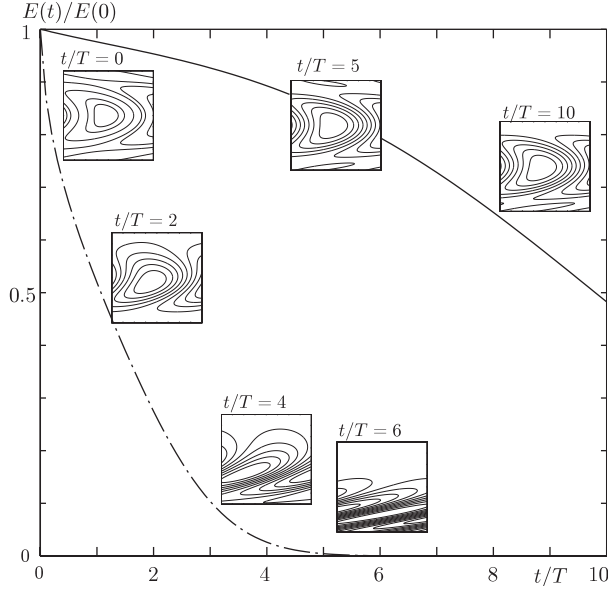}
\caption{The graphs depicts the evolution of the kinetic energy of
the viscous initial-value solution (\ref{eq:ViscousKelvinSolution})
at different Reynolds numbers and for shear rates $A = 1/T$. The dashed-dotted and
the solid lines depict the time-evolution of the energy for $\Rey =
10$ and $\Rey = 1000$, respectively. 
The insets show isolines of the real part of the stream function defined by
Eq. (\ref{eq:ViscousKelvinSolution}) at different points in
time in a moving frame of reference $\klamm{\tilde x,\tilde y}  \in
[-1,3]\times[-4,4]$ for $\tilde x = \kappa x - \frac{At^{2}}{2T}$
and $\tilde y = \kappa y - \frac{t}{T}$.} \label{fig:ViscousEnergy}
\end{center}
\end{figure}

In the viscous case, it is not possible to obtain \hilight{physically consistent} invariant solutions (see Appendix \ref{sec:AppendixViscousDivergingSolution}). 
Hence, we compute here the time-evolution of a configuration where the inviscid modes are imposed as initial conditions, but in a viscous setting.
For simplicity, we employ a superposition of Kelvin modes, weighted by $W_T$, to obtain
\begin{align}
\psi^{(V)}(x,y,t) =& \iint W_{T}\klamm{\kappa_{x},\kappa_{y}} \frac{\kappa_{x}^{2}+\kappa_{y}^{2}}{\kappa_{x}^{2} + \klamm{\kappa_{y} - \kappa_{x}At}^{2} }  e^{ i \kappa_{x}\klamm{ x-yAt} + i \kappa_{y}y } \times\notag\\
&\times\exp\klamm{-\nu t\left(\frac{1}{3}\kappa_x^2A^{2}t^2 -\kappa_y\kappa_x At +\kappa_y^2 + \kappa_x^2 \right)}  d\kappa_{x} d\kappa_{y}. \label{eq:ViscousKelvinSolution}
\end{align}
It can be readily seen that for finite times, contributions of high wave numbers $\kappa_{y} \to \infty$ will be quickly damped by viscous effects. As $W_{T}$ includes contributions from the complete range of wave numbers, we expect the shape and intensity of the modes to break down. In Figure \ref{fig:ViscousEnergy}, we contrast the longevity of the modes for a low-viscosity regime with the quite fast breakdown of the intensity and shape of the modes for higher viscosities.

\section{Conclusion \label{sec:Conclusion}}

We have presented a symmetry classification of the stream function form of
the linearized Navier-Stokes equation for two-dimensional perturbations. 
In particular, we have applied symmetry analysis to generate a 
general set of ansatz functions which goes beyond
the approaches known up-to-date, namely the normal mode approach and the Kelvin mode approach.
We found that for a general base flow, the equation allows for a time-
and space \hilight{translation} symmetry together with a scaling symmetry. If the
base flow is restricted to a linear shear flow, we obtain an
additional symmetry revealing specific properties of the flow.

We have shown that the normal mode approach as well as the Kelvin
mode approach can be systematically derived using successive
symmetry reductions.  The classical normal mode approach leading to
the Orr-Sommerfeld equation is based on the three symmetries of the
equation for a general base flow. Meanwhile, the Kelvin mode
approach is based on an additional symmetry obtained through the
restriction of the base flow to a linear shear flow. We note that
here, the complete set of symmetries is not used, rather the time
\hilight{translation} symmetry is excluded.

Including all relevant symmetries of the system leads to a new
invariant ansatz function exhibiting qualitatively different
behaviour. Kinematically, the new approach describes modes traveling
at a constant speed in the cross-stream direction and being
accelerated in the streamwise direction by the base flow. In the
inviscid case, we have presented an analytical closed-form solution
of these modes. The modes are energy-conserving in time and are
non-periodic/ decay  in the cross-stream direction. 
In the viscous case, the modes break down because of the contributions 
of high wave numbers $\kappa_y \to \infty$, which are quickly damped, 
in agreement with the expected behavior from the Kelvin modes.

We emphasise that the invariant approach presented in this work
is restricted to two-dimensional perturbations. Due to the constant
translation of the modes in the cross-stream direction, it is also
only applicable for finite times until the perturbations reach a boundary of the system.

\acknowledgements{
We are grateful to George Chagelishvili for reading various drafts of this paper and for giving valuable comments and suggestions on stability theory. We thank Alexei F. Cheviakov for various discussions on classifying the symmetries of the respective equations. We further thank the Center of Smart Interfaces (TU Darmstadt) for financial support through a seed fund project. Finally, we also thank Imperial College London for financial support through a DTG International Studentship.}

\appendix

\section{Viscous invariant Ansatz \label{sec:AppendixViscousDivergingSolution}}
We solve the linearized Navier-Stokes equation for a perturbation of a linear shear flow in stream function formulation (see Eq. (\ref{eq:StreamFunctionEqLin})) using the invariant ansatz function (\ref{eq:psis_S}). This approach leads to Eq. (\ref{eq:SelfSimilarModes_1}). Two naive solutions of this equation are
\begin{align}
g^{\HLbox{(\mathrm{I})}} _{1}\klamm{\tilde y} = e^{\tilde y}
\qquad \text{and} \qquad
g^{\HLbox{(\mathrm{I})}} _{2}\klamm{\tilde y} = e^{-\tilde y}.
\end{align}
The other two independent solutions of Eq. (\ref{eq:SelfSimilarModes_1}) are obtained by substituting
\begin{align}
u\klamm{\tilde y} = e^{\frac{\Rey}{2} \tilde y } \klamm{\frac{d^{2}}{d\tilde y^{2}} -1 } g\klamm{\tilde y}. \label{eq:Definition_uAppendix}
\end{align}
We then obtain the following second order ODE
\begin{align}
u \klamm{ 1 + \tilde c \Rey  + \klamm{\frac{\Rey}{2}}^{2}  + i \tilde y S \Rey  } - u'' = 0,
\end{align}
which is solved by the Airy-Functions
\begin{align}
u_{1}\klamm{\tilde y} &= \text{Ai}\klamm{  d_{1} \tilde y + d_{2}  }\\
\text{and}\qquad
u_{2}\klamm{\tilde y} &= \text{Bi}\klamm{ d_{1} \tilde y + d_{2}   }
\end{align}
with the parameters
\begin{align}
d_1 &\defi  \klamm{ i {S} \Rey}^{1/3} \\
\text{and} \qquad d_2 &\defi  \klamm{ i {S}\Rey  }^{-2/3}\klamm{1+\tilde c \Rey + \klamm{\frac{\Rey}{2}}^2}.
\end{align}
Inverting (\ref{eq:Definition_uAppendix}) yields
\begin{align}
g\klamm{\tilde y} &= \int_0^{\tilde y} \sinh\klamm{ {\tilde y} -  \tilde y'} e^{-\frac{\Rey}{2} \tilde y'} u(\tilde y')  d\tilde y',
\label{eq:Linear1_generalSol_2}
\end{align}
such that we obtain the independent solutions
\begin{align}
g^{\HLbox{(\mathrm{I})}} _{3}\klamm{\tilde y} =& \int_0^{\tilde y} \sinh\klamm{\tilde y -{\tilde y}'} e^{-\frac{\Rey}{2}{\tilde y}'} \text{Ai}(d_1 {\tilde y}' + d_2)d{\tilde y}',\\
g^{\HLbox{(\mathrm{I})}} _{4}\klamm{\tilde y} =& \int_0^{\tilde y} \sinh\klamm{\tilde y -{\tilde y}'} e^{-\frac{\Rey}{2}{\tilde y}'}  \text{Bi}\klamm{{ d_1{\tilde y}' + d_2 }}d{\tilde y}'. 
\end{align}
We note that all independent solutions $g_{1-4}$ diverge for $\tilde y \to \pm \infty$, violating the consistency of the initial condition.

%


\end{document}